\documentclass[graybox]{svmult}
\usepackage{type1cm}
\usepackage{graphicx}
\graphicspath{{Images/}}
\usepackage{pgfplots}
\usepackage{multicol}
\usepackage[bottom]{footmisc}
\usepackage{newtxtext}
\usepackage{newtxmath} 
\makeindex 

\usepackage{fancyvrb}

\usepackage[nameinlink]{cleveref}
\Crefname{lstlisting}{listing}{listings}
\crefname{algocf}{algorithm}{algorithms}

\newcommand{\specialcell}[2][c]{%
\begin{tabular}[#1]{@{}c@{}}#2\end{tabular}}

\definecolor{HighlightColor}{rgb}{0.43, 0.69, 0.26}
\definecolor{HighlightColor2}{rgb}{0.945, 0.349, 0.373}
\definecolor{HighlightColor3}{rgb}{0.349, 0.604, 0.827}
\definecolor{HighlightColor4}{rgb}{0.976, 0.651, 0.353}
\definecolor{HighlightColor5}{rgb}{.62, 0.4, 0.671}
\definecolor{HighlightColor6}{rgb}{.804, 0.439, 0.345}
\definecolor{HighlightColor7}{rgb}{.843, 0.498, 0.702}
\definecolor{HighlightColor8}{rgb}{0.745, 0.769, 0.349}
\definecolor{HighlightColor9}{rgb}{0.25, 0.25, 0.25}
\definecolor{HighlightColor10}{rgb}{0.5, 0.5, 0.5}
\definecolor{CodeBG}{rgb}{0.95,0.95,0.95}

\usepackage{listings}
\usepackage[ruled]{algorithm2e}
\usepackage{siunitx}
\usepackage{tikz}
\usetikzlibrary{calc}

\makeatletter
\pgfkeys{/eye/.cd,
	x/.code           = {\def\eye@x{#1}},
	y/.code           = {\def\eye@y{#1}},
	rotation/.code    = {\def\eye@rot{#1}},
	radius/.code      = {\def\eye@rad{#1}}
}

\newcommand{\eye}[1][]{%
\pgfkeys{/eye/.cd,
	x         = 0,
	y         = 0,
	rotation  = 0,
	radius    = 1
}
\pgfqkeys{/eye}{#1}
	\draw[rotate around={\eye@rot:(\eye@x,\eye@y)}]
		(\eye@x,\eye@y) -- ++(-.5*55:.9*\eye@rad)
		(\eye@x,\eye@y) -- ++(.5*55:.9*\eye@rad);
	\draw[fill=white] (\eye@x,\eye@y) ++(\eye@rot+27.5:.75*\eye@rad) arc (\eye@rot+27.5:\eye@rot-27.5:.75*\eye@rad) -- (\eye@x, \eye@y) -- cycle;
	\draw[gray,shorten >= .5, shorten <= .3] ($(\eye@x,\eye@y) + (\eye@rot+55/3:.75*\eye@rad)$) -- ($(\eye@x,\eye@y) + (\eye@rot-55/3:.75*\eye@rad)$);
	\draw[fill=gray] (\eye@x,\eye@y) ++(\eye@rot+55/3:.75*\eye@rad) arc (\eye@rot+180-55:\eye@rot+180+55:.28*\eye@rad);
	\draw[fill=gray] (\eye@x,\eye@y) ++(\eye@rot+55/3:.75*\eye@rad) arc (\eye@rot+55/3:\eye@rot-55/3:.75*\eye@rad);
	\fill[rotate around={\eye@rot:(\eye@x,\eye@y)}] (\eye@x,\eye@y) ++(.723*\eye@rad, 0) circle (0.035 and 0.1);
}
\makeatother

\begin{document}
\title*{Massively Parallel Path Space Filtering}
\author{Nikolaus Binder, Sascha Fricke, and Alexander Keller}
\institute{Nikolaus Binder (NVIDIA), Sascha Fricke (University of Braunschweig), and Alexander Keller (NVIDIA)
}

\maketitle

\abstract*{copy of below}

\abstract{
	Restricting path tracing to a small number of paths per pixel for performance reasons
	rarely achieves a satisfactory image quality for scenes of interest.
	However, path space filtering may dramatically improve the visual quality
	by sharing information across vertices of paths classified as proximate.
	Unlike screen space-based approaches, these paths neither need to be present on the screen,
	nor is filtering restricted to the first intersection with the scene.
	While searching proximate vertices had been more expensive than filtering in screen space,
	we greatly improve over this performance penalty
	by storing, updating, and looking up the required information in a hash table.
	The keys are constructed from jittered and quantized information,
	such that only a single query very likely replaces costly neighborhood searches.
	A massively parallel implementation of the algorithm is demonstrated on a graphics processing unit (GPU).
}

\section{Introduction}
\label{Sec:Introduction}

\begin{figure}[ht]
	\centering
	\begin{tikzpicture}
		\node[anchor = south west, inner sep = 0, outer sep = 0] at (0, 0) {%
			\includegraphics[width=.8\linewidth, trim = 0 150 0 200, clip]{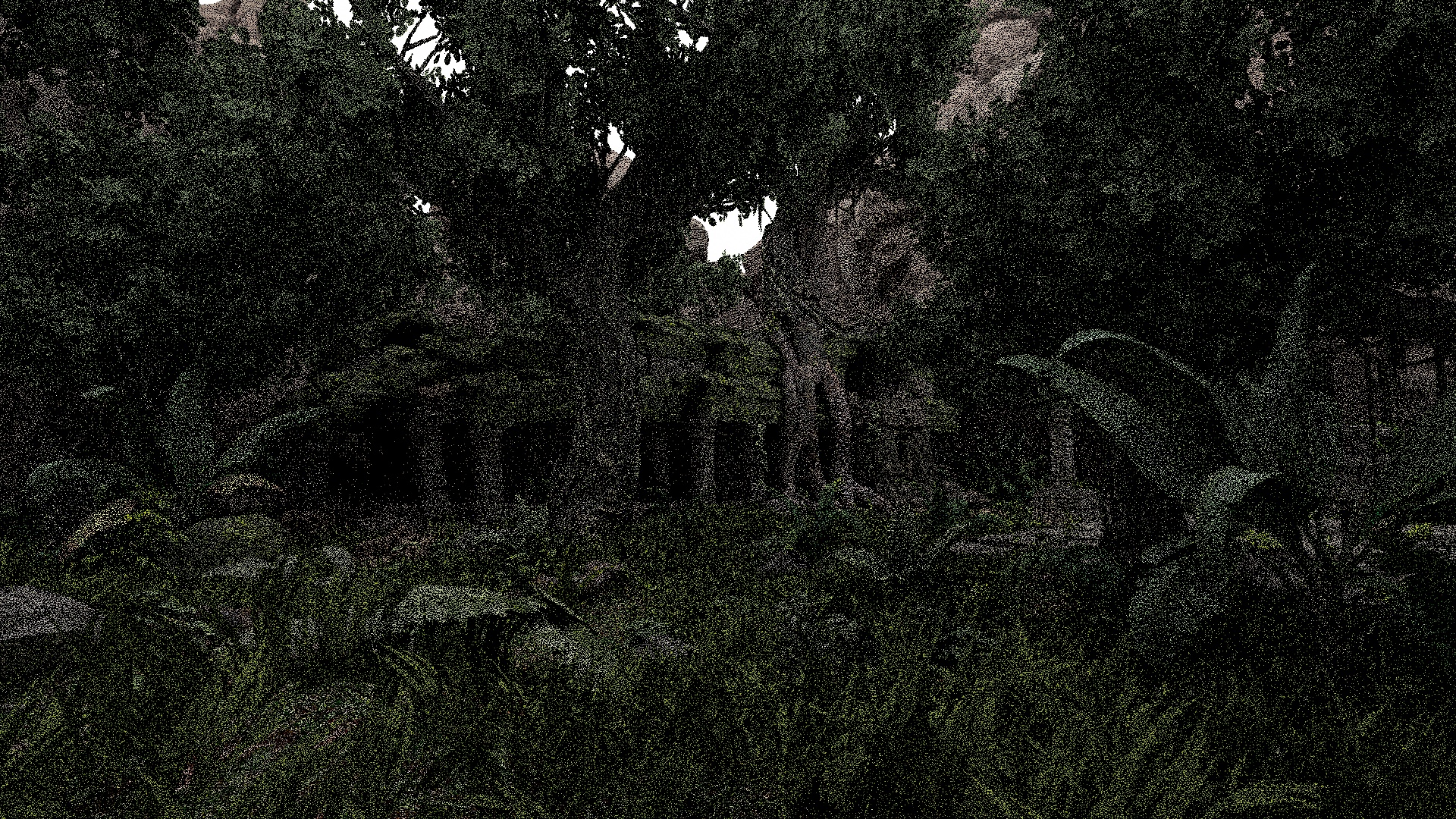}%
			\includegraphics[width=.1995\linewidth, trim = 400 423 1400 474, clip]{TempleEntranceFromJungleLeft_pt__textured__1spp}%
		};
		\draw[thick, HighlightColor2] (1.95, 1.4) rectangle ++(0.625, 0.85);
	\end{tikzpicture}\\[.5em]
	\begin{tikzpicture}
		\node[anchor = south west, inner sep = 0, outer sep = 0] at (0, 0) {%
			\includegraphics[width=.8\linewidth, trim = 0 150 0 200, clip]{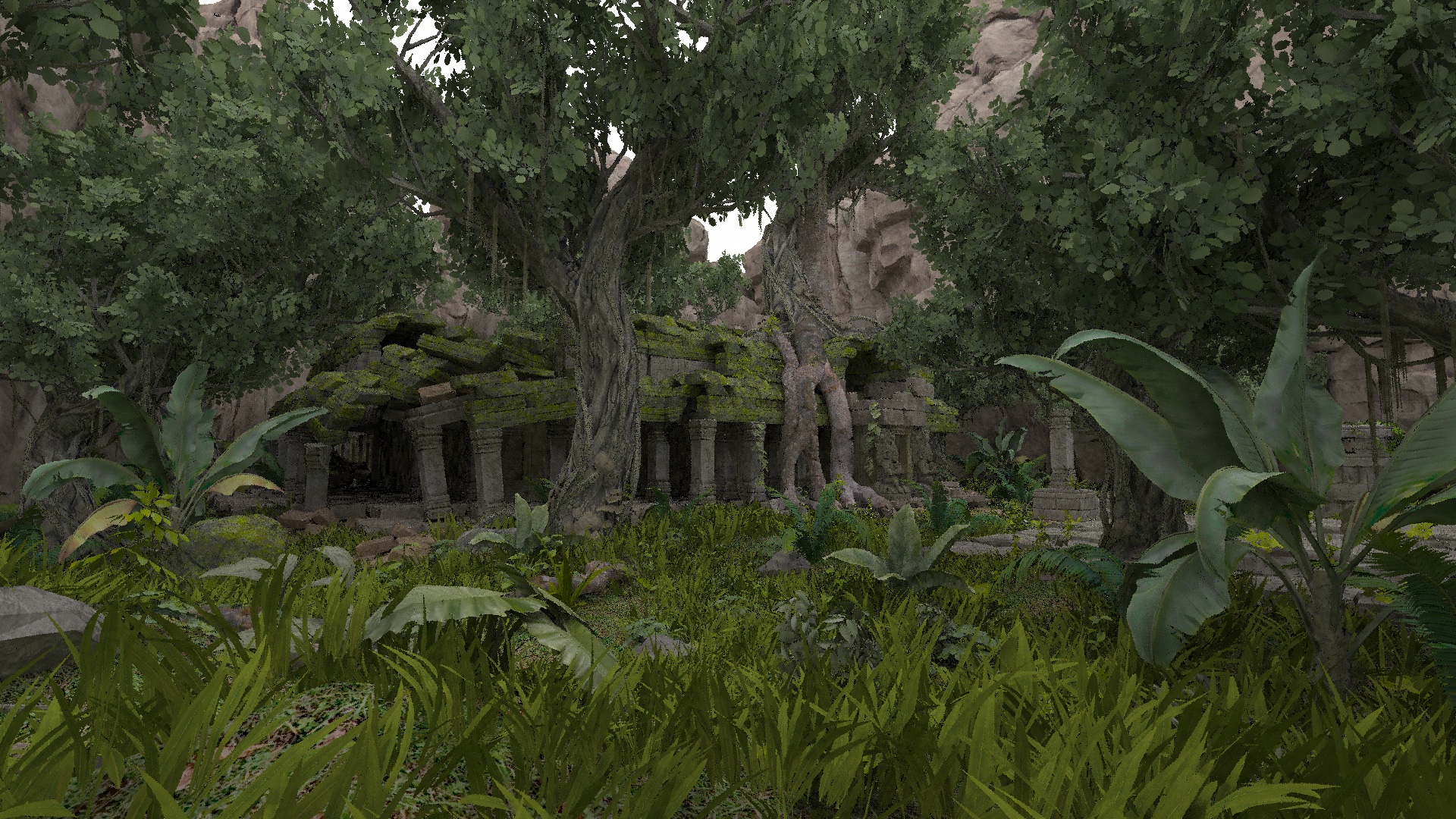}%
			\includegraphics[width=.1995\linewidth, trim = 400 423 1400 474, clip]{TempleEntranceFromJungleLeft_hprogr__textured__1spp}%
		};
		\draw[thick, HighlightColor2] (1.95, 1.4) rectangle ++(0.625, 0.85);
	\end{tikzpicture}\\[.5em]
	\begin{tikzpicture}
		\node[anchor = south west, inner sep = 0, outer sep = 0] at (0, 0) {%
			\includegraphics[width=.8\linewidth, trim = 0 150 0 200, clip]{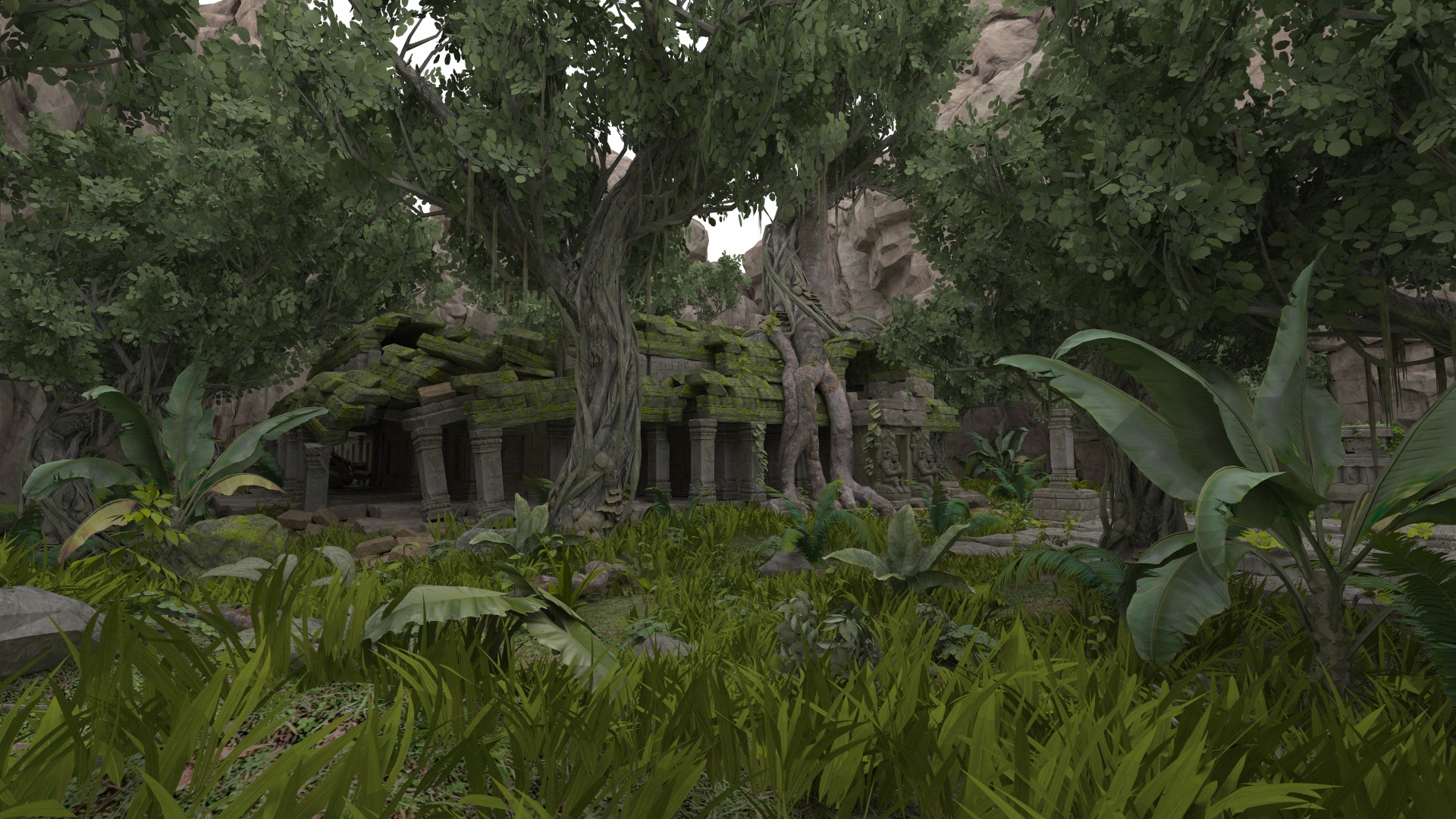}%
			\includegraphics[width=.1995\linewidth, trim = 400 423 1400 474, clip]{TempleEntranceFromJungleLeft_pt__textured__1024spp}%
		};
		\draw[thick, HighlightColor2] (1.95, 1.4) rectangle ++(0.625, 0.85);
	\end{tikzpicture}\\[.5em]
	\caption{
		Path tracing at one path per pixel (top)
		in combination with hashed path space filtering (middle)
		very closely approximates the reference solution using 1024 paths per pixel (bottom)
		and does so with an overhead of about \SI{1.5}{ms} in HD resolution.
		Scene courtesy of Epic Games.
  }
	\label{Fig:Rainforest}
\end{figure}

Realistic image synthesis consists of high-dimensional numerical integration of functions with potentially high
variance. Restricting the number of samples therefore often results in visible noise, which efficiently can be reduced
by path space filtering \cite{PathSpaceFiltering} as shown in
 \Cref{Fig:Rainforest}.

We improve the performance of path space filtering by replacing costly neighborhood search with averages of clusters
in voxels resulting from quantization. Our new algorithm is suitable for interactive and even real-time rendering and it enables
many applications trading a controllable bias for a dramatic speedup and noise reduction.

\begin{figure}
	\footnotesize
	\tikzset{every picture/.style={line width=0.7pt}}
	\begin{tabular*}{\linewidth}{@{\extracolsep{\fill}} cc}
		\begin{tikzpicture}[scale=1.58]
			\begin{scope}[scale=0.5]
				\eye[x=-0.5, y=1.5, rotation=45];
			\end{scope}
			\draw[-latex] (0, 1) -- (0.72, 1.87) -- (1.37, 0.65);
			\draw (0.17, 1.68) -- (1.55, 2.2);
			\draw (0.09, 1.34) -- (0.35, 0.98) node[font=\small, anchor=north west, inner sep=0, outer sep=1] {$P$};
			\draw (0.85, 0.65) -- (1.9, 0.65);

			\draw (1.7, 1.85) -- (2.42, 1.67);

			\draw (2.05, 0.3) -- (2.75, 0.48);
			\draw (2.4, 2.02) rectangle (3.09, 2.19);
				\node[anchor=north, inner sep=0, outer sep=4, font=\small] at (1.37, 0.65) {$x$};
			\node[anchor=west, inner sep=0, outer sep=4, font=\small] at (3.09, 2.105) {$L_e$};
				\draw[densely dashed] (1.37, 0.65) -- (1.37, 1.35);
				\draw[densely dashed] (1.37, 0.65) ++ (57:.6) arc (57:90:.6);
				\node[anchor=south west, font=\small, shift={(-0.07, 0.5)}] at (1.37, 0.65) {$\theta_x$};
				\draw[densely dashed, -latex] (1.37, 0.65) -- (2.07, 1.75) -- (2.4, 0.42) -- (2.92, 2.02);
				\draw[densely dashed] (1.37, 0.65) -- (2.07, 1.75);
				\draw[densely dashed] (2.07, 1.75) -- ++(-0.17*1.15, -0.72*1.15);
				\draw[densely dashed] (2.07, 1.75) ++ (-104:.8) arc (-104:-122:.8);
				\node[anchor=north east, font=\small, shift={(-0.12, -0.72)}] at (2.07, 1.75) {$\theta_y$};
				\node[anchor=south, inner sep=0, outer sep=4, font=\small, shift={(0.05,0)}] at (2.07, 1.75) {$y$};
		\end{tikzpicture}&
		\begin{tikzpicture}[scale=1.58]
			\begin{scope}[scale=0.5]
				\eye[x=-0.5, y=1.5, rotation=45];
			\end{scope}
			\draw[-latex] (0, 1) -- (0.72, 1.87) -- (1.37, 0.65);
			\draw (0.17, 1.68) -- (1.55, 2.2);
			\draw (0.09, 1.34) -- (0.35, 0.98) node[font=\small, anchor=north west, inner sep=0, outer sep=1] {$P$};
			\draw (0.85, 0.65) -- (1.9, 0.65);

			\draw (1.7, 1.85) -- (2.42, 1.67);

			\draw (2.05, 0.3) -- (2.75, 0.48);
			\draw (2.4, 2.02) rectangle (3.09, 2.19);
				\node[anchor=north, inner sep=0, outer sep=4, font=\small] at (1.32, 0.65) {$x$};
				\node[anchor=north, inner sep=0, outer sep=4, font=\small] at (1.45, 0.705) {$x'$};
			\node[anchor=west, inner sep=0, outer sep=4, font=\small] at (3.09, 2.105) {$L_e$};
				\draw[densely dashed] (1.45, 0.65) -- (1.45, 1.35);
				\draw[densely dashed] (1.45, 0.65) ++ (60:.6) arc (60:90:.6);
				\node[anchor=south west, font=\small, shift={(-0.09, 0.5)}] at (1.45, 0.65) {$\theta_{x'}$};
				\draw[densely dashed] (2.07, 1.75) -- ++(-0.17*1.15, -0.72*1.15);
				\draw[densely dashed] (2.07, 1.75) ++ (-104:.9) arc (-104:-120:.9);
				\node[anchor=north east, font=\small, shift={(-0.15, -0.90)}] at (2.07, 1.75) {$\theta_y$};
				\node[anchor=south, inner sep=0, outer sep=4, font=\small, shift={(0.05,0)}] at (2.07, 1.75) {$y$};
				\draw[densely dashed] (1.37, 0.65) circle (0.3);
				\draw[densely dashed] (0, 1) -- (0.95, 1.96) -- (1.45, 0.65);
				\draw[latex-latex] (1.45, 0.65) -- (2.07, 1.75) -- (2.4, 0.42) -- (2.92, 2.02);
		\end{tikzpicture}\\
		\specialcell{(a) forward path tracing, subpath\\ connections, and next event estimation} & \specialcell{(b) density estimation and\\path space filtering}
	\end{tabular*}
	\caption{Geometry of light transport simulation by path tracing.
	Starting from the eye on the left through the image plane $P$,
	a light transport path segment ends in $x$.
	(a) Forward path tracing continues the path until the path terminates on the surface of a light source.
	A path can also be completed by directly connecting to a vertex $y$ on the surface of a light source (next event estimation)
	or to a vertex $y$ of the same or a different path (subpath connection).
	(b) Radiance in point $x$ can directly be evaluated by accumulating radiance in vertices $x'$ in a local neighborhood
	either for density estimation or path space filtering.
	}
	\label{Fig:LightTransport}
\end{figure}
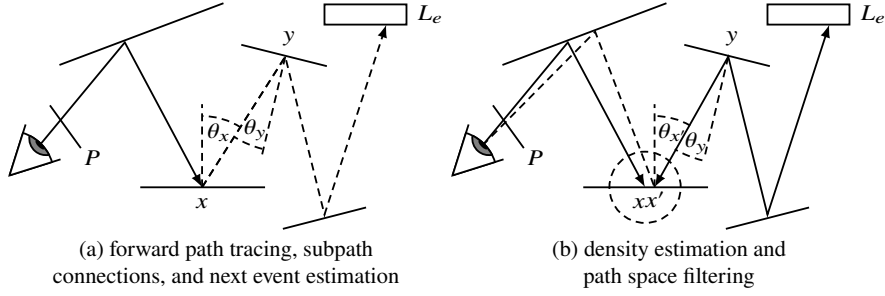

\section{Light Transport Simulation}
\label{Sec:LightTransportSimulation}

Light transport is simulated by tracing rays
that connect the surfaces of the light sources and the camera sensor through a three-dimensional scene,
represented by surfaces, materials, and scattering properties of the materials and in volume.
Upon interaction with the surfaces or participating media in the volume, a new segment of the path is generated.
The connection between adjacent path segments is called vertex,
and all information of its path and the interaction is referred to as its descriptor.
While in reality paths of photons start on emissive surfaces,
and the camera sensor measures the ones terminating on its surface,
simulations most often construct paths backwards.
Somewhat counter-intuitively, a simulation in reverse photon direction is called ``forward path tracing'' --
forward in view direction.

At each point $x$ and direction $\omega$, the incident radiance $L_i$ is the sum of the emitted radiance
$L_e$ and the reflected radiance $L_r$ in the point $x$ and in direction $\omega$:
\begin{equation}
	L_i(x, \omega) = L_e(x, \omega) + L_r(x, \omega)\label{Eqn:IncidentRadiance}
\end{equation}
Given the distance $d := \left\lVert c - x \right\rVert_2$ between the center of the ball $c$ and a point $x$,
the characteristic function of the ball $B$ with radius $r$ is defined as
\begin{align}
	\chi_B(d, r) &:=
	\begin{cases}
		1 & d^2 < r^2\\
		0 & \textrm{otherwise.}
	\end{cases}
	\label{Eqn:CharFBall}
\end{align}
The following \crefrange{Eqn:SolidAngle}{Eqn:PSF} show four different ways to formulate the reflected radiance $L_r$ in a point $x$ in direction $\omega_r$
and \cref{Fig:LightTransport} illustrates the resulting techniques:
\begin{alignat}{2}
	\lefteqn{ L_r(x, \omega_r)} \nonumber \\
	& = \int_{{\mathcal S}^2_-(x)}%
	L_i(x, \omega) f_r(\omega_r, x, \omega) \cos \theta_{x} d\omega \label{Eqn:SolidAngle} \\
	& = {\int_{\partial V} V(x, y)} L_i(x, \omega) f_r(\omega_r, x, \omega) \cos \theta_x \frac{\cos \theta_{y}}{|x - y|^2} dy \label{Eqn:Surface} \\
	& =
	\begin{aligned}
	\lim_{r(x) \rightarrow 0} \int_{\partial V} \int_{{\mathcal S}^2_-(y)} & \frac{\chi_B\bigl(x - h(y, \omega), r(x)\bigr)}{\pi r(x)^2} L_i\bigl(h(y, \omega), \omega\bigr) \cdot\\
	& \cdot f_r\bigl(\omega_r, h(y, \omega), \omega\bigr) \cos \theta_y d\omega dy
	\end{aligned}
	\label{Eqn:PMap}\\
	& = \lim_{r(x) \rightarrow 0} \int_{{\mathcal S}^2_-(x)} \frac{\int_{\partial V} \chi_B\bigl(x - x', r(x)\bigr) w(x, x') L_i(x', \omega) f_r(\omega_r, x, \omega) \cos \theta_{x'} dx'}{\int_{\partial V} \chi_{B}\bigl(x - x', r(x)\bigr)w(x, x') dx'} d\omega\label{Eqn:PSF}
\end{alignat}

\begin{description}
\item[\textbf{Forward Path Tracing:}]
		\Cref{Eqn:SolidAngle} integrates radiance over the upper hemisphere $S_{-}^2$
		by multiplying the incident radiance $L_i$ from the angle $\omega$ in the point $x$
		with the spatio-directional reflectivity $f_r$ for the two angles in the point $x$
		and the cosine of the incident ray to account for the change of area of the projected solid angle.
		Inserting the equation into \cref{Eqn:IncidentRadiance} and the result back into \cref{Eqn:SolidAngle}
		allows for subsequently prolonging paths and is known as \emph{(recursive) forward path tracing}.
	\item[\textbf{Next Event Estimation and Subpath Connection:}]
		\Cref{Eqn:Surface} changes the integration domain to the scene surface $\partial V$
		so that a path can be constructed in which the point $x$ connects to any point $y$ on the scene surfaces.
		The integrand is then extended by the mutual visibility $V$ of the two points $x$ and $y$.
		The fraction of the cosine of the second angle and the squared distance of the two points accounts for the change of measure.
		One often refers to this fraction as the \emph{geometric term}.
		\Cref{Eqn:Surface} is especially useful since it allows to directly connect to the surface of a light source (\emph{next event estimation})
		or any other vertex of any path (\emph{subpath connection}).
	\item[\textbf{Density Estimation:}]
		\Cref{Eqn:PMap} again integrates over the scene surface.
		However, it realizes density estimation by restricting to a local neighborhood
		in a sphere with radius $r(x)$ using the characteristic function of the ball $\chi_B$.
		The density is then obtained by dividing by the area of the circle that stems from the intersection of the ball and the flat surface.
		For a radius going to zero, the formulation is equivalent.
		Density estimation tracing photons from the light sources is usually referred to as \emph{photon mapping}.
	\item[\textbf{Path Space Filtering:}]
		\Cref{Eqn:PSF} similarly integrates over a local neighborhood of $x$.
		In contrast to density estimation, a local weighted average is calculated.
		This local average is normalized by the integral of all weights in the neighborhood instead of the area of a circle.
		While for a sphere with vanishing radius this formulation is again equivalent,
		the method trades a certain bias for a reduction of variance for any non-zero radius.
		Due to the filtering of the local average this technique has been introduced as \emph{path space filtering} \cite{PathSpaceFiltering}.
\end{description}
While implementations of \crefrange{Eqn:SolidAngle}{Eqn:PSF} each individually come with their own strengths and weaknesses
and are therefore often combined for robustness in offline rendering applications~\cite{VeachPhD,LafortuneBDPT,GeorgievPhd},
time constraints of real-time image synthesis as well as advances in \emph{path guiding}~\cite{LightRL,MuellerPathGuiding,MuellerNeuralGuiding}
lead to the vast majority of implementations only employing \cref{Eqn:SolidAngle,Eqn:Surface}.
Our work aims at a substantial acceleration of path space filtering so that it can be used in real-time applications, too, resulting in a considerable variance reduction.

\subsection{Previous Work}

Filtering results of light transport simulation is gaining more and more attention in real-time, interactive, and even offline rendering.
The surveys by Zwicker et al. \cite{Zwicker:2015} and Sen et al. \cite{SiggraphFiltering} present an overview of recent developments.
The fastest approaches use only information available at primary intersections
and perform filtering in screen space.
Further recent work is based on deep neural networks \cite{Bako:2017,Chaitanya:2017},
hierarchical filtering with weights based on estimated variance in screen space \cite{Schied:2017},
or on improving performance by simplifying the overall procedure \cite{Mara:2017}.

Path space filtering~\cite{PathSpaceFiltering}, on the other hand, averages contributions of light transport paths in path space,
which allows for filtering at non-primary intersections
and for a more efficient handling of dis-occlusions during temporal filtering.
However, querying the contributions in path space so far had been significantly more expensive than filtering the contributions of neighboring pixels in screen space.
Neglecting the fact that locations that are close in path space are not necessarily adjacent in screen space
enables interactive filtering in screen space~\cite{GautronPSF}.
As a consequence, filtering is almost only a good approximation for primary rays or reflections from sufficiently smooth and flat surfaces.
In fact, such filtering algorithms are a variant of a bilateral filtering using path space proximity to determine weights.
Sharing information across pixels according to a similarity measure dates back as early as the 1990s~\cite{KellerPhd}.
Since then, several variants have been introduced, for example by re-using paths in nearby pixels~\cite{BekaertReusing},
for filtering by anisotropic diffusion~\cite{MccoolDiffusion}
or using edge-avoiding \'A-Trous wavelets~\cite{DammertzPhd}.

Fast filtering is also possible in texture space~\cite{TextureSpaceFiltering},
which at least requires a bijection between the scene surface and texture space.
While this may be tricky already, issues may arise along discontinuities of a parametrization in addition.
Furthermore, only filtering is restricted to locations on surfaces when operating in texture space,
and thus volumetric effects must be filtered separately.

Hachisuka et al. also use a hash table in a light transport simulation on the GPU~\cite{HachisukaPMapGPU}.
The approach is fundamentally different in two aspects:
First, it traces photons and stores them in the hash table for density estimation,
while our method averages radiance in vertices from arbitrary light paths.
Second, their method implements simple sampling with replacement in voxels,
culling all but one photon per voxel. Our method does the exact opposite:
It collects radiance from all paths whose vertices coincide in a voxel.

Mara et al. summarize and evaluate a number of methods for photon mapping on the GPU~\cite{MaraPMapGPU}.
Their evaluation also includes work from Ma and McCool using hash tables with lists of photons in per voxel~\cite{MaMcCoolPMap}.
While all examined methods may be used for path space filtering instead of photon mapping,
their performance is at least limited by the maintenance of lists.

Havran et al. use two trees for final gathering with photon mapping~\cite{HavranFG}.
The overhead of tree construction and traversal
as well as iterating through lists of vertices severely limit the performance in our intended real-time use case.

Multiple Importance Sampling weights, for example those for path space filtering, can be further optimized \cite{ContinuousMIS}.

Kontkanen et al. explore irradiance filtering, a subset of path space filtering~\cite{IrradFiltering}.
Spatial caching of shading results in a hash table for walkthroughs of static scenes~\cite{DietrichCaching}
uses similar methods to the ones presented in this work for the lookup of these results.
Again, the method can be seen as a subset of Path Space Filtering:
It is restricted to caching diffuse illumination
and neither includes filtering
nor spatial and temporal integration in an arbitrary number of vertices of a light path.

\section{Algorithm}
\label{Sec:Algorithm}

Similar to path space filtering~\cite{PathSpaceFiltering},
the input of the method is a set of vertices in which radiance should be filtered.
In addition, the method receives information from their paths, such as attenuation and throughput.
Therefore, the overall simulation must run in at least two phases:
The first phase generates paths and stores all necessary information,
similar to a path tracing simulation without filtering.

After a set of light transport paths has been generated,
the second phase of the algorithm filters radiance in voxels, as described in \cref{Sec:AveragingInVoxels},
and stores as well as looks up the averages in a hash table, see \cref{Sec:HashMap}.
Techniques described in \cref{Sec:HandlingLowDensity} improve the robustness of the algorithm in voxels with a small number of vertices.
\Cref{Sec:InteractiveSimulations} discusses additional filtering opportunities and challenges in interactive light transport simulations.

Finally, after filtering,
for each selected vertex its associated average is multiplied by its throughput
and accumulated in its respective pixel.
The throughput is the attenuation from the camera along the light transport path up to the selected vertex.

\subsection{Averaging in Voxels}
\label{Sec:AveragingInVoxels}

The new algorithm builds upon a different characteristic function to be used in \cref{Eqn:PSF}.
Unlike the characteristic function of a three-dimensional ball in \cref{Eqn:CharFBall},
it uses the key $k$ constructed from the descriptor of a vertex instead of only depending on its position $x$.

\Cref{Sec:CharFVoxel} defines this characteristic function,
and \cref{Sec:SelectionOfComponents} describes the construction of a key $k$ from the descriptor of a vertex in detail.
\Cref{Sec:AdaptiveResolution} and \cref{Sec:Jittering} present extensions to the method that reduce the perceived bias.

\subsubsection{Characteristic Function of a Voxel}
\label{Sec:CharFVoxel}
Given a resolution selection function $s(k)$, the characteristic function of the voxel is defined as
\begin{equation}
	\chi_V(k, k') :=
	\begin{cases}
		1 & \lfloor s(k)k \rfloor = \lfloor s(k')k' \rfloor \wedge s(k) = s(k')\\
		0 & \textrm{otherwise.}
	\end{cases}
	\label{Eqn:CharacteristicFunctionVoxel}
\end{equation}
In the simplest case, $s(k)$ is a constant which defines the size of all voxels.
In practice, other heuristics improve the quality of the approximation.
For example, it is often advisable to increase the world space extent of a voxel with its distance to the camera sensor along the path.
Then, one can filter more aggressively in distant voxels,
and increasing the size counteracts the decrease of the density of vertices from paths directly coming from the camera sensor with increasing distance.
Our implementation parameterizes $s(k)$ by defining an area on the screen,
and then calculates the projected size of the area on the screen using the projection theorem.

\begin{figure}[htbp]
	\footnotesize
	\begin{tabular*}{\linewidth}{@{\extracolsep{\fill}} cc}
		\includegraphics[width=.45\textwidth]{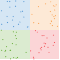}&
		\includegraphics[width=.45\textwidth]{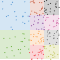}\\
		(a) $s(k) \propto 1$ & (b) $s(k) \propto 2^{\lfloor \log_2 x_0 \rfloor}$
	\end{tabular*}
	\caption{The characteristic function of a voxel $\chi_V(k, k')$ is 1 for all $k, k'$ inside the same voxel, here depicted with the same color for (a) constant resolution selection function $s(k)$ resulting in uniform voxel size or (b) increasing $s(k)$ from left to right, which subsequently shrinks the voxels.}
	\label{Fig:CharacteristicFunctionVoxel}
\end{figure}

The transitivity of the characteristic function of the voxel,
i.e $\chi_V(k, k') = \chi_V(k, k'') \Leftrightarrow \chi_V(k, k') = \chi_V(k', k'')$,
implies that the set of path vertices can be partitioned into disjoint sets:
$\chi_V(k, k')$ equals one for any two vertices with keys $k, k'$ of the same set, and is always zero otherwise.
\Cref{Fig:CharacteristicFunctionVoxel} shows examples for sets of two-dimensional keys defined by this characteristic function.
In practice, keys are at least three-dimensional, i.e. defined by the world space position of the vertex.
\Cref{Sec:SelectionOfComponents} explains the selection of components of the key in detail.

Replacing $\chi_B$ in \cref{Eqn:PSF} with $\chi_V$, setting $w(x, x') := 1$,
and using the result as an approximation instead of only considering its limit, we get
\begin{equation}
	L_r(x, \omega_r)
	\approx
	\int_{{\mathcal S}^2_-(x)} \frac{\int_{\partial V} \chi_V(k, k') L_i(x', \omega) f_r(\omega_r, x, \omega) \cos \theta_{x'} dx'}{\int_{\partial V} \chi_V(k, k') dx'} d\omega.\label{Eqn:VoxelPSF1}
\end{equation}
If $f_r(\omega_r, x, \omega)$ is separable into $f_r(\omega_r, x) \cdot f_i(x, \omega)$,
and $f_i(x, \omega)$ is -- at least approximately -- constant within the voxel,
we will be able to rewrite and rearrange \cref{Eqn:VoxelPSF1} to become
\begin{equation}
L_r(x, \omega_r)
\approx
f_r(\omega_r, x) \int_{{\mathcal S}^2_-(x)} \frac{\int_{\partial V} \chi_V(k, k') L_i(x', \omega) f_i(x', \omega) \cos \theta_{x'} dx'}{\int_{\partial V} \chi_V(k, k') dx'} d\omega.\label{Eqn:VoxelPSF}
\end{equation}
In the following, we call the product $L_i(x', \omega) \cdot f_i(\omega_r, x', \omega) \cdot \cos\theta_{x'}$ the \emph{contribution} of the vertex in $x'$.
All terms of the integrand except for $\chi_V(k, k')$ are independent of $x$ and $\omega_r$,
and $\chi_V(k, k')$ is identical for all vertices with key $k$ in a set.
Therefore it is now possible to calculate the integral in \cref{Eqn:VoxelPSF} only once for the set of vertices in each voxel.

\subsubsection{Construction of Keys}
\label{Sec:SelectionOfComponents}

The key $k$ of a vertex contains a subset of the components of its descriptor.
Thereby, the integral calculated per voxel is independent of those components excluded from the key.
The selection of components of the key is critically important for defining the tradeoff between bias and variance reduction:
Including additional components of the descriptor in the key may reduce the bias,
while excluding components allows for the inclusion of more vertices in the integral, therefore reducing variance.
In the following we will give a short overview of components of the descriptor that one would typically include in the key.

The quality of the approximation in \cref{Eqn:VoxelPSF} highly depends on the deviation of $L_i$ in $x'$ from the one in $x$.
First and foremost, it is therefore recommended to restrict the world space extent of the voxel
by including the position $x$ of the vertex in the key $k$.
While $L_i$ is not continuous in practice -- for example in edges of sharp shadows --
the perceived error of approximation decreases with the world space extent of a voxel.

In practice, one can furthermore not guarantee that $\lim_{x' \to x}\cos\theta_{x'} = \cos\theta_{x}$
due to different surface orientations in the two locations, for example in sharp edges of objects.
Including the normal of the surface in the point $x$ in the key avoids the resulting ``smearing'' across edges and ``flattening'' of surfaces.
On the other hand, the quantization of the normal by \cref{Eqn:CharacteristicFunctionVoxel} causes discontinuities in smooth normals.
\Cref{Sec:Jittering} details how one can make the artifacts less perceptually pronounced and more amenable to additional filtering.

Splitting $f_r(\omega_r, x, \omega)$ into $f_r(\omega_r, x) \cdot f_i(x, \omega)$ is not always possible.
On highly reflective surfaces, $f_r$ is defined as a Dirac delta function, and filtering is pointless.
Therefore, vertices on such surfaces should not be selected in the first place.
On glossy surfaces, however, filtering may reduce variance efficiently, again at the cost of a certain bias.
While $f_r$ can not be split on these surfaces without unpleasantly and undesirably changing the visual appearance,
splitting the domain of incident angles and computing separate averages for each interval may be a viable tradeoff.
Therefore, for vertices on such surfaces one can append the incident angle to the key
so that the quantization of \cref{Eqn:CharacteristicFunctionVoxel} splits the domain of the incident angle, too.

Materials are often composed of different layers with different properties.
Then, filtering the layers independently offers the opportunity to use different resolutions $s(k)$
as well as constructing keys with different components for the different layers.
For example, a material consisting of a glossy layer on top of a diffuse one
could only include the angle $\omega_r$ in the key used for filtering the glossy layer
since the attenuation of the diffuse layer is independent of it.
Thus, the integral used for the diffuse layer benefits from including more samples.
Appending an identifier of the layer to the key splits the average into several individual ones, which can be combined later.

Note that depending on the number of selected components for the key,
voxels are not necessarily three-dimensional,
and their extent may vary between components.

\subsubsection{Adaptive Resolution}
\label{Sec:AdaptiveResolution}

\begin{figure} \includegraphics[width=\linewidth, trim = 0 150 0 150, clip]{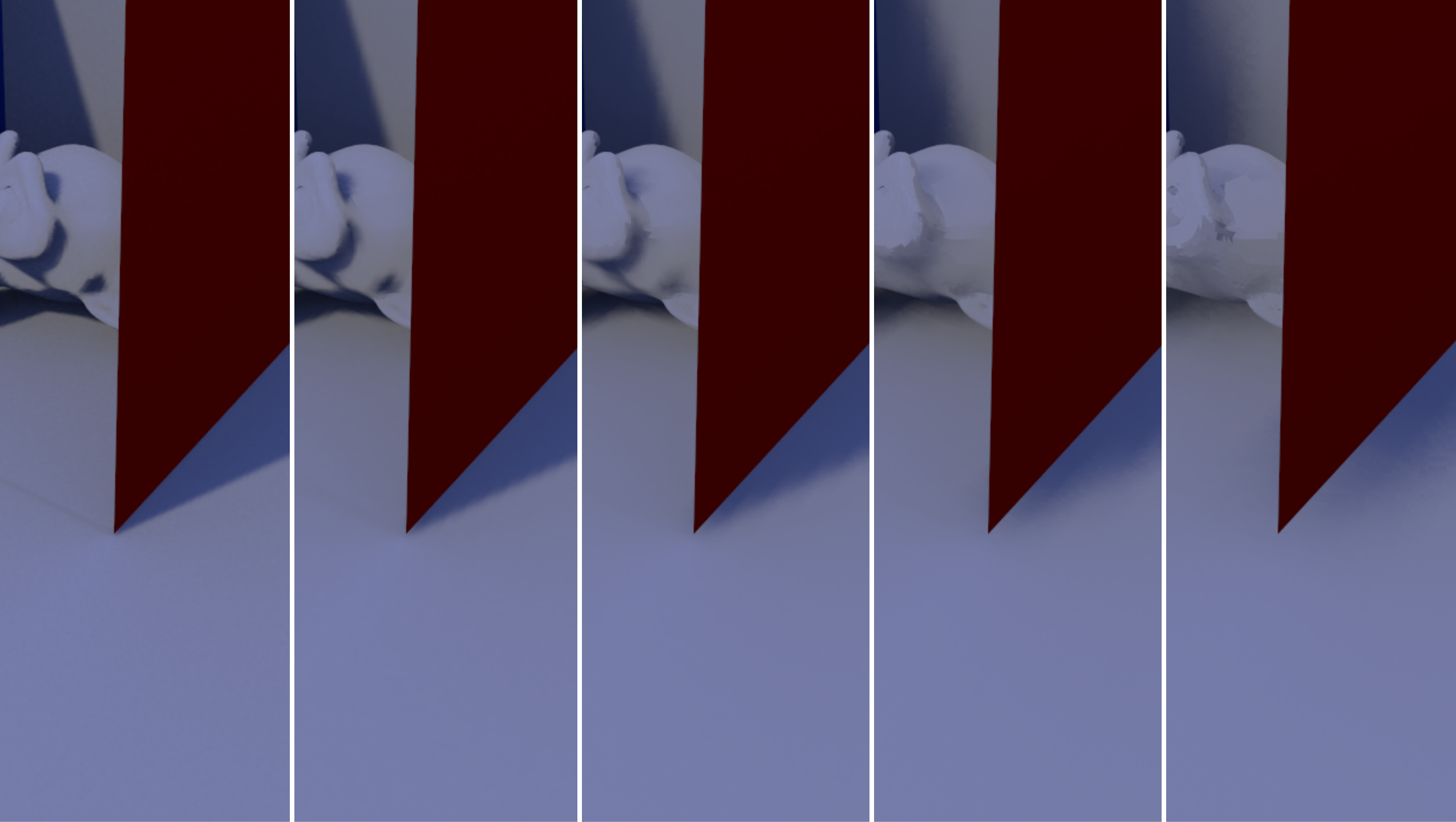}
	\caption{Increasing the filter size by lowering the resolution $s(k)$ (left to right) increasingly blurs shadows and
		also increases the amount of light and shadow leaking.}
	\label{Fig:LightShadowLeaks}
\end{figure}

The choice of the resolution selection function $s(k)$ is crucial for finding a good tradeoff between perceived bias and reduction of variance.
In theory, its value should be large in areas with a lot of high frequency detail in $L_i$.
Unfortunately, those areas are almost always unknown in practice since $L_i$ is unknown.
\Cref{Fig:LightShadowLeaks} shows how a sharp shadow is blurred due to averaging radiance in a large voxel.
One would therefore like to adaptively chose a finer resolution along its boundary.

Finite spatial differences may be used in heuristics for adaptation.
While their computation either introduces a certain overhead or reduces the number of independent samples,
cost may be amortized over frames in environments changing only slowly over time.
Note that finite differences only estimate spatial variations of the averages,
and one must therefore carefully both choose and adjust such heuristics
as well as determine the number samples used for finite spatial differences.

\subsubsection{Filter Kernel Approximation by Jittering}
\label{Sec:Jittering}

\begin{figure}
	\includegraphics[width=.42\textwidth]{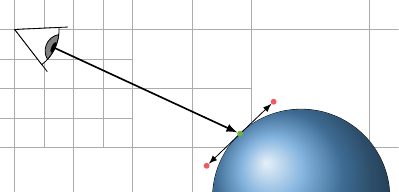} \hfill
	\includegraphics[width=.28\linewidth]{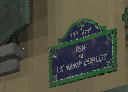} \hfill
	\includegraphics[width=.28\linewidth]{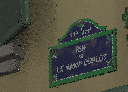}
	\caption{Jittering trades quantization artifacts for noise. Left: Note that the
		resolution $s(k)$ at the jittered location (red) may differ from the one of the original
		location (green). Spatial jittering hides otherwise visible
		quantization artifacts (middle): The resulting noise (right) is more amenable to
		the eye and much simpler to remove by a secondary filter.}
	\label{Fig:Jittering}
\end{figure}

\begin{figure}[htbp]
	\begin{tabular*}{\linewidth}{@{\extracolsep{\fill}} cccc}
		\includegraphics[width=0.24\linewidth, trim={0 108 0 108}, clip]{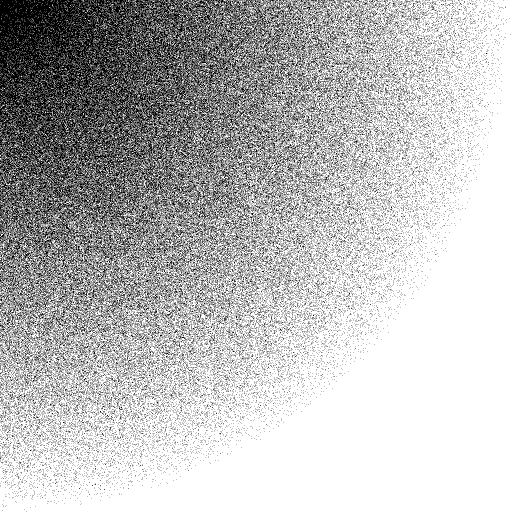}%
		& \includegraphics[width=0.24\linewidth, trim={0 108 0 108}, clip]{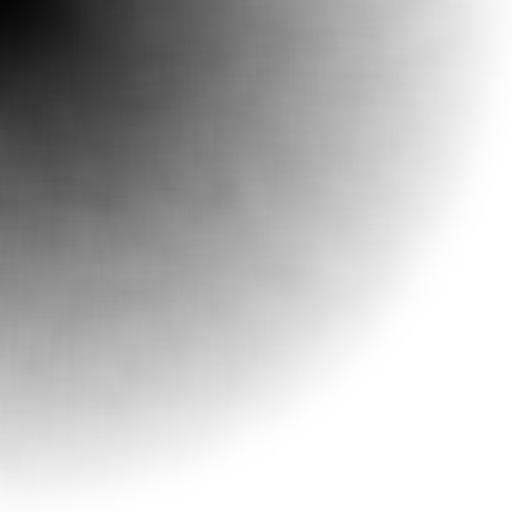}%
		& \includegraphics[width=0.24\linewidth, trim={0 108 0 108}, clip]{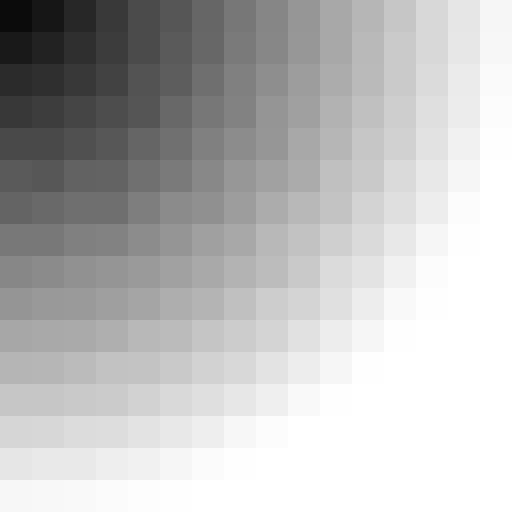}%
		& \includegraphics[width=0.24\linewidth, trim={0 108 0 108}, clip]{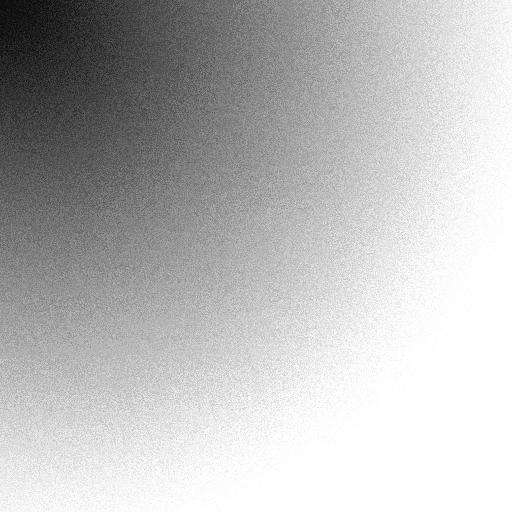}\\
		(a)%
		& (b)%
		& (c)%
		& (d)%
	\end{tabular*}
	\caption{
		Two-dimensional example for averages in voxels:
		The noisy input (a)
		is filtered in each vertex by path space filtering (b).
		Instead, the new method filters in each voxel, resulting in block artifacts (c).
		Additionally jittering before accumulation and lookup resolves the artifacts in noise.
	}
	\label{Fig:Example2d}
\end{figure}

The discontinuities of quantization are removed by jittering components of the key,
which in fact amounts to approximating a filter kernel by sampling.
Jittering depends on the kind of component of the key,
for example, positions are jittered in the tangent plane of an intersection, see \cref{Alg:Hash}.
The resulting noise is clearly preferable over the visible discretization artifacts resulting from quantization,
as illustrated and shown in \cref{Fig:Jittering,Fig:Example2d}.
In contrast to discretization artifacts, noise from jitter is simple to filter.

\subsection{Accumulation and Lookup in a Hash Table}
\label{Sec:HashMap}

The averages in each voxel can be calculated in two different ways:
First, each voxel can gather radiance of all included vertices.
This process may run in parallel over all voxels and does not require any synchronization.
On the other hand, a list of voxels as well as a list of vertices per voxel must be maintained.
The second way to calculate the average radiance in a voxel runs in parallel over all vertices:
Each vertex atomically adds its contribution $L_i(x', \omega)f_i(x', \omega)\cos\theta_{x'}$ to a running sum of the voxel
and increments the counter of the voxel.
The average is finally calculated by dividing the sum by the counter.
While the latter approach requires atomic operations, it does not involve maintenance of any lists.
Furthermore, summation can be parallelized over the paths or over the vertices,
matching the parallelization scheme of typical light transport simulations.
Finally, parallelization per path or per vertex exposes more parallelism,
and therefore the second approach outperforms the first one on modern \emph{graphics processing units} (GPUs) significantly.

Accumulation with the latter approach needs a mapping from the key of a vertex to the voxel with its running sum and counter.
Typically, the set of voxels is sparse since vertices are mostly on two-dimensional surfaces in three-dimensional space.
Additional components of the key increase sparsity even further.

Hash tables provide such a mapping in constant time for typical sets of keys:
First, a hash of the key is calculated using a fast hash function.
A modulo operator then wraps this hash into the index range of the table cells.
Since both the hash function as well as the modulo operator are not bijective,
different keys may be mapped to the same index.
Therefore, an additional check for equality of keys is required,
and keys must also be stored in the table.
\Cref{Sec:Fingerprinting} details a cheaper alternative for long keys.

Upon index collision with a different key, linear probing subsequently increments the index,
checking if the table cell at the updated index is empty or occupied with an entry with same key.
Then, the collision has been resolved.
There exist various other collision resolution methods that improve upon several aspects of linear probing
and have proven to be more efficient in certain use cases, especially for hash tables with high occupancy.
On the other hand, we do not primarily aim to minimize the size of the hash table,
and our experiments show that linear probing comes with a negligible overhead if the table is sufficiently large.
We restrict the number of steps taken for linear probing to avoid performance penalties of extreme outliers,
and resort to the unfiltered contribution of the vertex if the number of steps exceeds this limit.
So far, our choices for the table size and number of steps so extremely rarely resulted in such failures
that further improvement has been deemed unnecessary.
\Cref{Sec:SearchingByLinearProbing} broadens the application of linear probing
from only collision resolution to an additional search for similar voxels.

\subsubsection{Fingerprinting}
\label{Sec:Fingerprinting}
Instead of storing and comparing the rather long keys for checking equality of keys,
we calculate a shorter fingerprint \cite{RabinFingerprinting, LS-Hashing} from a second, different hash function of the same key
and use it for this purpose, see \cref{Alg:Hash}.
Excluding a sentinel value from the fingerprint, we can furthermore use this sentinel to mark empty cells.

Using fingerprints instead of the full keys is a tradeoff between correctness and performance:
In theory, fingerprints of different keys may coincide.
In practice, our choice of \SI{32}{bit} fingerprints never caused any collision in our evaluation of several test scenes and numerous simulations.
Still, there is a certain probability of failure,
and we deliberately favor the tiny probability of a failure over the performance penalty of storing and comparing full keys.

\begin{algorithm}
	\DontPrintSemicolon
	\SetKwFunction{hash}{hash}
	\SetKwFunction{hashtwo}{hash2}
	\SetKwFunction{verify}{verify}
	\SetKwFunction{rand}{jitter}
	\SetKwFunction{slod}{level\_of\_detail}
	\KwIn{Location $x$ of the vertex, the normal $n$, the position of the camera $p_{\textrm{cam}}$, and the scale $s$. }
	\KwOut{Hash $i$ to determine the position in the hash table and hash $f$ for fingerprinting.}
	$l \gets$ \slod{$|p_{\textrm{cam}} - x|$}\;
	$x' \gets x +$ \rand{$n$}$\ \cdot\ s \cdot 2^l$\;
	$l' \gets$ \slod{$|p_{\textrm{cam}} - x'|$}\;
	$\tilde{x} \gets \left\lfloor \frac{x'}{s \cdot 2^{l'}}\right\rfloor$\;
	$i \gets$ \hash{$\tilde{x}, \ldots$}\;
	$f \gets$ \hashtwo{$\tilde{x}, n, \ldots$}\;
 
	\caption{
		Computation of the two hashes used for lookup.
		Note that the arguments of a hash function, which form the key,
		may be extended to refine clustering.
	}
	\label{Alg:Hash}
\end{algorithm}

\subsubsection{Searching by Linear Probing}
\label{Sec:SearchingByLinearProbing}
\begin{figure}
	\centering
	\begin{tikzpicture}[scale=0.7, rotate=-90]
		\fill[HighlightColor] (-2, -5.5) rectangle ++(1, 1);
		\draw[thick, HighlightColor3, -latex] (-1.5, -5.1) -- ++(-1/0.82, 0);
		\draw[thick, HighlightColor4, -latex] (-1.6, -4.7) -- ++(-1, 0.7);
		\draw[thick, HighlightColor2, -latex] (-1.2, -5.3) -- ++(-1, -0.7);

		\node at (0.5, -11) {a)};
		\node at (2.5, -11) {b)};
		\node at (4.5, -11) {c)};

		\foreach \i in {0,...,9}
		{
			\pgfmathparse{Mod(\i,2)==0?1:0}
			\ifnum\pgfmathresult>0
				\fill[opacity=0.15] (0, -\i) rectangle ++(1, -1);
			\else
				\fill[opacity=0.2] (0, -\i) rectangle ++(1, -1);
			\fi
			\ifthenelse{\i=1} {\fill[HighlightColor2, opacity=1] (0, -\i) rectangle ++(1, -1);}{}
			\ifthenelse{\i=6} {\fill[HighlightColor4, opacity=1] (0, -\i) rectangle ++(1, -1);}{}
			\ifthenelse{\i=9} {\fill[HighlightColor3, opacity=1] (0, -\i) rectangle ++(1, -1);}{}
		}

		\foreach \i in {0,...,9}
		{
			\pgfmathparse{Mod(\i,2)==0?1:0}
			\ifnum\pgfmathresult>0
				\fill[opacity=0.15] (2, -\i) rectangle ++(1, -1);
			\else
				\fill[opacity=0.2] (2, -\i) rectangle ++(1, -1);
			\fi
		}
		\fill[HighlightColor2, opacity=1] (2, -9) rectangle ++(1, -0.15);
		\fill[HighlightColor2, opacity=1] (2, -9.85) rectangle++(1, -0.15);
		\fill[HighlightColor2, opacity=1] (2, -9.15) rectangle ++(0.15, -0.7);
		\fill[HighlightColor2, opacity=1] (2.85, -9.15) rectangle ++(0.15, -0.7);

		\fill[HighlightColor3, opacity=1] (2.15, -9.15) rectangle ++(0.7, -0.15);
		\fill[HighlightColor3, opacity=1] (2.15, -9.7) rectangle ++(0.7, -0.15);
		\fill[HighlightColor3, opacity=1] (2.15, -9.3) rectangle ++(0.15, -0.4);
		\fill[HighlightColor3, opacity=1] (2.7, -9.3) rectangle ++(0.15, -0.4);

		\fill[HighlightColor4, opacity=1] (2.3, -9.3) rectangle ++(0.4, -0.4);

		\foreach \i in {0,...,9}
		{
			\pgfmathparse{Mod(\i,2)==0?1:0}
			\ifnum\pgfmathresult>0
				\fill[opacity=0.15] (4, -\i) rectangle ++(1, -1);
			\else
				\fill[opacity=0.2] (4, -\i) rectangle ++(1, -1);
			\fi
			\ifthenelse{\i=9} {\fill[HighlightColor2, opacity=1] (4, -\i) rectangle ++(1, -1);}{}
			\ifthenelse{\i=7} {\fill[HighlightColor4, opacity=1] (4, -\i) rectangle ++(1, -1);}{}
			\ifthenelse{\i=8} {\fill[HighlightColor3, opacity=1] (4, -\i) rectangle ++(1, -1);}{}
		}
	\end{tikzpicture}
	\caption{Instead of including normals in the key a) to differentiate
	contributions whose vertices fall into the same voxel b), the fingerprint of
	a key may include normal information. This allows
	one to differentiate normal information by linear probing as shown in c).}
	\label{Fig:LinearProbing}
\end{figure}
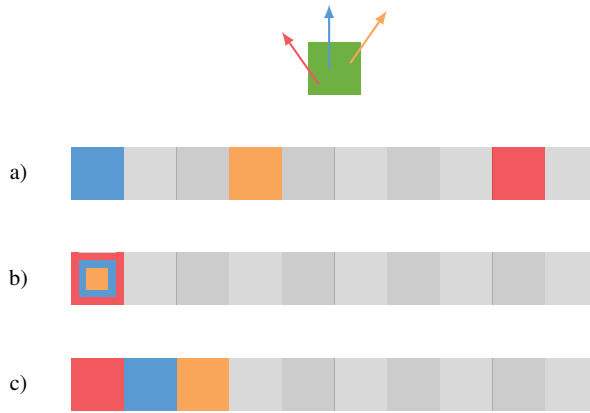

In addition, linear probing may be used to differentiate attributes of the light transport path
at a finer resolution as shown in \Cref{Fig:LinearProbing}:
For example, normal information may be included in the key handed to the fingerprinting hash function
instead of already including it in the main key.
This allows one to search for similar normals by linear probing.

Note that due to other voxels also possibly occupying neighboring cells in the hash table,
searching with linear probing must go beyond mismatching fingerprints.
Therefore, the method works best if both the number of additional contributions
as well as the occupancy of the hash table are low.

\subsection{Handling Voxels with a low Number of Vertices}
\label{Sec:HandlingLowDensity}

Often, there exists a tiny number of voxels that only contain very few vertices,
and therefore variance reduction in those voxels is suboptimal.
Examples of such voxels include those that are only slightly overlapped by objects.
\Cref{Sec:NeighborhoodSearch} and \cref{Sec:MultiLevel} present two approaches
that avoid high variance in voxels with only very few vertices
at an almost negligible overhead.

\subsubsection{Neighborhood Search}
\label{Sec:NeighborhoodSearch}

Accumulation in voxels by using quantized keys and a hash table requires a single \texttt{atomicAdd} operation per component of the radiance of each vertex,
and the final average is computed with one additional non-atomic read operations per component for the sum and one for the counter.
So, as long as access to the hash table happens in constant time,
calculation of the average also takes constant time.

This is in sharp contrast to existing methods computing sums or averages in a spatial neighborhood
which for each vertex take linear time for a search within a given radius
or (typically) logarithmic time for a fixed number of neighboring vertices.

\begin{figure}
	\centering
	\includegraphics[width=.5\textwidth]{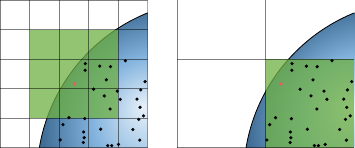}
	\caption{
		Instead of searching the $\mathbf{n^s}$ neighborhood in $s$ dimensions (left),
		we utilize clustering resulting from quantization at a lower resolution to accumulate contributions,
		which allows for a single look up (right).
		We only resort to an additional neighborhood search
		in the rare case that the number of vertices in a voxel falls below a certain threshold.
		}
	\label{Fig:Cluster}
\end{figure}

Even if primarily only one average per voxel is computed and looked up,
searching for neighboring voxels can still be valuable:
In theory, increasing the resolution $s(k)$
and additionally searching for neighboring voxels may result in variance reduction similar to the one at a lower resolution,
however then with a lower bias.
Yet, the number of neighbors grows exponentially with the number of components of the key.
Thus, such an approach is currently ruled out by the time constraints for real-time applications.
\Cref{Fig:Cluster} shows a comparison between neighborhood search and clustering by selecting a coarser resolution.

As a fallback, neighborhood search is still very valuable:
If the number of vertices in a voxel falls below a certain threshold,
we allow for an additional search.
We observe that given an appropriate threshold,
the number of such voxels is so low that the overhead is completely negligible
while the perceptual improvement is clearly visible.

\subsubsection{Multi-level Accumulation}
\label{Sec:MultiLevel}

Special treatment of voxels with averages from only very few vertices is important for visual fidelity:
Even if the number of voxels with a high variance is very low,
it may be very visible, especially since their appearance is so different from the rest.
Besides searching in a local neighborhood to reduce variance in these cases (see \cref{Sec:NeighborhoodSearch}),
selecting a coarser resolution also effectively increases the number of vertices in the local average --
at the price of an increased bias.
Using more than one resolution at a time avoids the chicken-and-egg problem that arises from first selecting an appropriate resolution,
and then, after accumulation according to this resolution,
determining that it has been set too high or too low.

For simulations in interactive scenarios,
one may also select the resolution based on information from previous frames,
see \cref{Sec:TemporalAccumulation}.

\subsection{Accumulation over Time}
\label{Sec:InteractiveSimulations}

Reusing contributions across frames dramatically increases efficiency.
However, attention should be paid to arising pitfalls and efficiency aspects:
\Cref{Sec:TemporalAccumulation} details the differences and similarities between filtering and integration across frames,
\cref{Sec:TemporalResolution} explains handling of resolution changes across frames,
and \cref{Sec:Eviction} outlines how the amount of information kept over frames can be limited
to avoid running out of available memory in the hash table.

\subsubsection{Temporal Filtering and Temporal Integration}
\label{Sec:TemporalAccumulation}

For static scenes, the averages will converge.
For dynamic environments, maintaining two sets of averaged contributions
and combining them with an exponential moving average
$c = \alpha \cdot c_{\textrm{old}} + (1 - \alpha) \cdot c_{\textrm{new}}$
is a common tradeoff between convergence and temporal adaptivity.

However, combining the averages $c_{\textrm{old}}$ and $c_{\textrm{new}}$ by an exponential moving average
is not equivalent to temporal integration.
Especially averages in voxels with relatively few samples do not converge.
In fact, setting $\alpha := \frac{N_{\textrm{old}}}{N_{\textrm{old}} + N_{\textrm{new}}}$ correctly integrates across frames.
On the other hand, temporal integration is only possible
if the underlying setting, including lighting conditions and object positions,
remains unchanged across frames.

A first, simple heuristic is to
accumulate samples over time up to a certain degree.
This may be implemented using a fixed threshold for the number of samples
and accumulating samples across frames until reaching it.
Note that the heuristic is completely unaware of changes in the scene.

\begin{figure}[htbp]
	\begin{tabular*}{\linewidth}{@{\extracolsep{\fill}} ccc}
		\includegraphics[width=.32\textwidth, trim={300px 200px 1000px 400px}, clip]{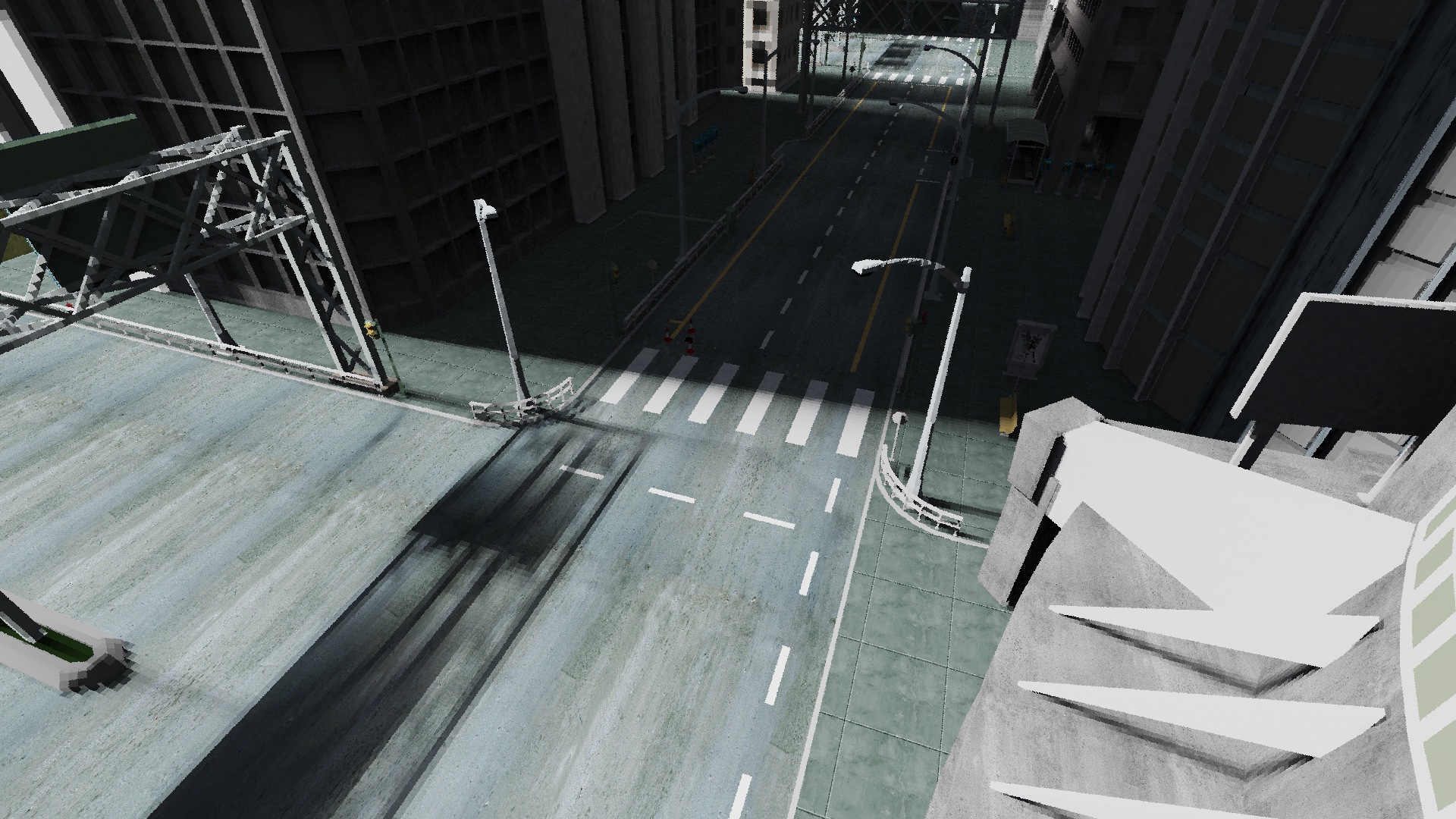}&
		\includegraphics[width=.32\textwidth, trim={300px 200px 1000px 400px}, clip]{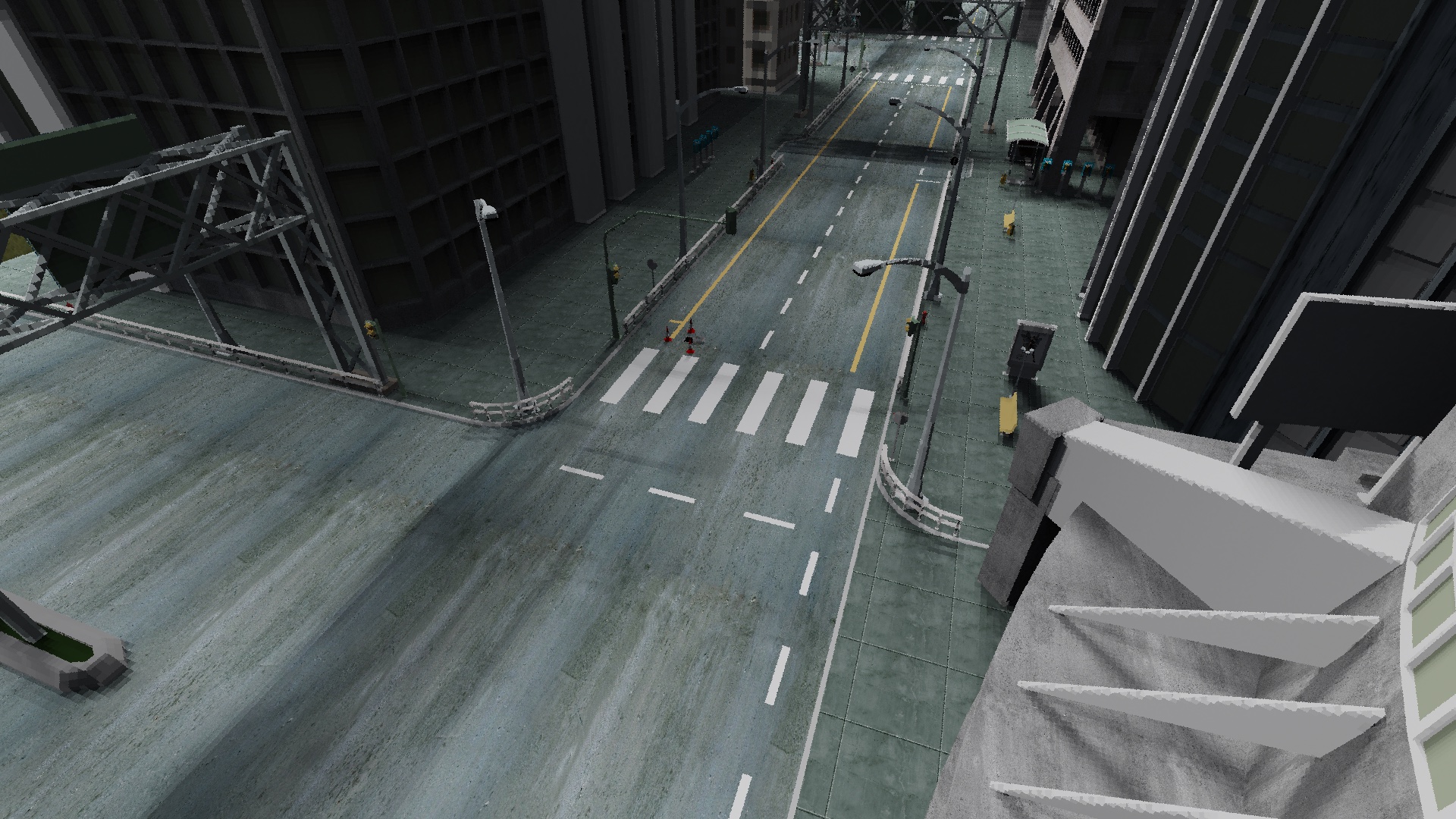}&
		\includegraphics[width=.32\textwidth, trim={300px 200px 1000px 400px}, clip]{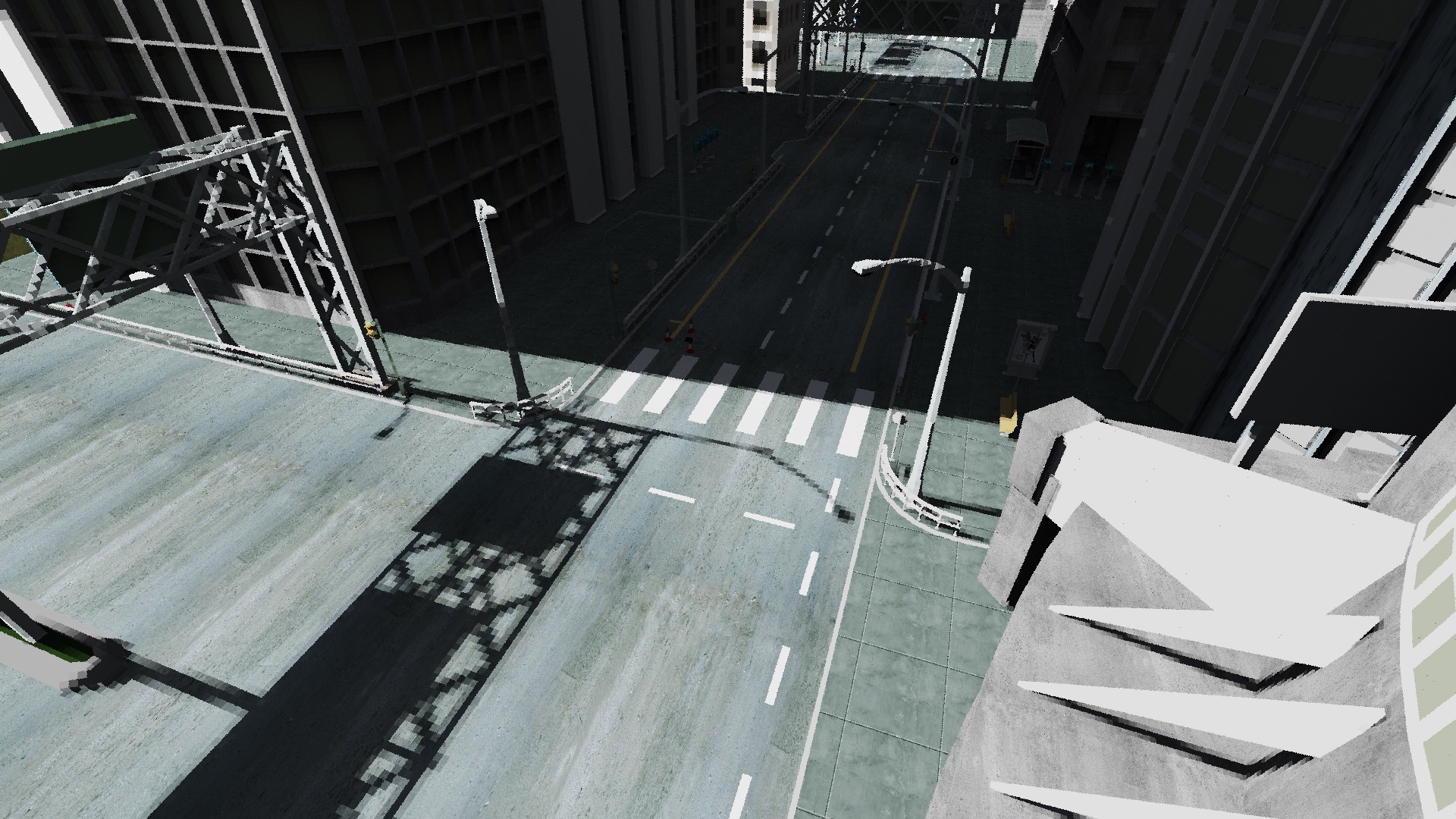}\\
		(a) temporal filtering&
		(b) temporal integration&
		(c) hybrid
	\end{tabular*}
	\caption{
		In a scene with a moving light source,
		temporal filtering using an exponential moving average blurs shadow boundaries (a),
		and temporal integration averages out the entire shadow (b).
		Adaptively blending $\alpha$ between zero (for large temporal differences)
		and $\alpha := \frac{N_{\textrm{old}}}{N_{\textrm{old}} + N_{\textrm{new}}}$ (for no temporal differences)
		combines both and preserves sharp shadow boundaries (c).
		Note that jittering has deliberately been disabled for all to emphasize sharp boundaries further.
	}
  \label{Fig:TempFilteringVsTempIntegration}
\end{figure}

A second, more expensive heuristic builds upon temporal finite differences:
A number of paths is re-evaluated with the same parameters,
and the difference of their contribution to the original ones allows to detect changes that affect the current voxel.
Similar to the spatial finite differences in \cref{Sec:AdaptiveResolution},
the additional cost may be amortized over frames,
and the number of samples used for finite differences
as well as their influence on the balance between temporal adaptation and temporal integration
must be carefully optimized.
Note that averaging in voxels can be used for the samples used for finite differences, too.
\Cref{Fig:TempFilteringVsTempIntegration} shows a comparison between temporal filtering, temporal integration
and a hybrid blending both based on temporal finite differences.
A similar approach for screen space filtering has been explored in detail by Schied et al.~\cite{SchiedSVGF2}.

\subsubsection{Changes in Resolution across Frames}
\label{Sec:TemporalResolution}

In many simulations, the camera is dynamic,
and therefore the resolution of a voxel may change across frames
if it depends on the position of the camera.
Then, already collected contributions in a voxel at one resolution must be copied to a voxel at either a higher or lower resolution.

If the resolution in the new frame decreases,
one can simply add up the contributions in voxels of higher resolution.
Since we store sums and counters, both only need to be added individually.

If the resolution in the new frame increases,
the contributions in voxels of lower resolution must be distributed to voxels at a higher resolution.
Due to the lack of resolution, this case is much nuanced:
On the one hand, using already collected contributions lowers variance,
but on the other hand, the coarser resolution may become unpleasantly visible.
One therefore needs to find a good compromise between the two,
and set $\alpha$ accordingly.

Finite spatial or temporal differences can also be filtered across frames, requiring attention in similar aspects.

\subsubsection{Eviction Strategy}
\label{Sec:Eviction}

Evicting contributions of voxels which have not been queried for a certain period of time is necessary for larger scenes and changing camera.
Besides the least recently used (LRU) eviction strategy,
heuristics based on longer term observations are efficient.

A very simple implementation relies on replacing the most significant bits of the fingerprinting hash
by a priority composed of for example the number of vertices in the voxel and last access time during temporal filtering.
Thus the pseudo-randomly hashed least significant bits guarantee eviction to be uniformly distributed across the scene,
while the most significant bits ensure that contributions are evicted according to priority.
This allows collision handling and eviction to be realized by a single \texttt{atomicMin} operation.

\section{Results and Discussion}
\label{Sec:Results}

While filtering contributions at primary intersections with the proposed algorithm is quite fast,
it only removes some artifacts of filtering in screen space.
However, hashed path space filtering has been designed to target real-time light transport simulation:
It is the only efficient fallback when screen space filtering fails or is not available,
for example, for specular or transparent objects.

Filtering on non-diffuse surfaces requires to include additional parameters in the key
and heuristics such as increasing the quantization in areas with non-diffuse materials to minimize the visible artifacts.

Filtering, and especially accumulating contributions, is always prone to light and shadow leaking (see \cref{Fig:LightShadowLeaks}),
which is the price we pay for performance.
Some artifacts may be ameliorated by employing suitable heuristics as reviewed in \cite[Sec.2.1]{PathSpaceFiltering} and in \cref{Sec:AdaptiveResolution}.

The new algorithm filters incoherent intersections at HD resolution ($1920 \times 1080$ pixels)
in about \SI{3}{ms} on an NVIDIA Titan V GPU.
Filtering primary intersections doubles the performance due to the more coherent memory access patterns.

The image quality is determined by the filter size,
which balances noise versus blur as shown in \Cref{Fig:LightShadowLeaks}.
Both the number of collisions in the hash table
and hence the performance of filtering
also depend on the size of the voxels.
We found specifying the voxel size by $s$-times the projected size of a pixel most convenient.
Note that maximum performance does not necessarily coincide with best image quality.
The hash table size is chosen proportional to the number of pixels at target resolution
such that potentially one vertex could be stored per pixel.
In practice, filtering requires multiple vertices to coincide in a voxel,
and therefore the occupancy of the hash table is rather low.
Such a small occupancy improves the performance
as it lowers the number of collisions and time spent for collision resolution.

\begin{figure}
	\centering
	\includegraphics[width=\linewidth, trim = 0 150 0 100, clip]{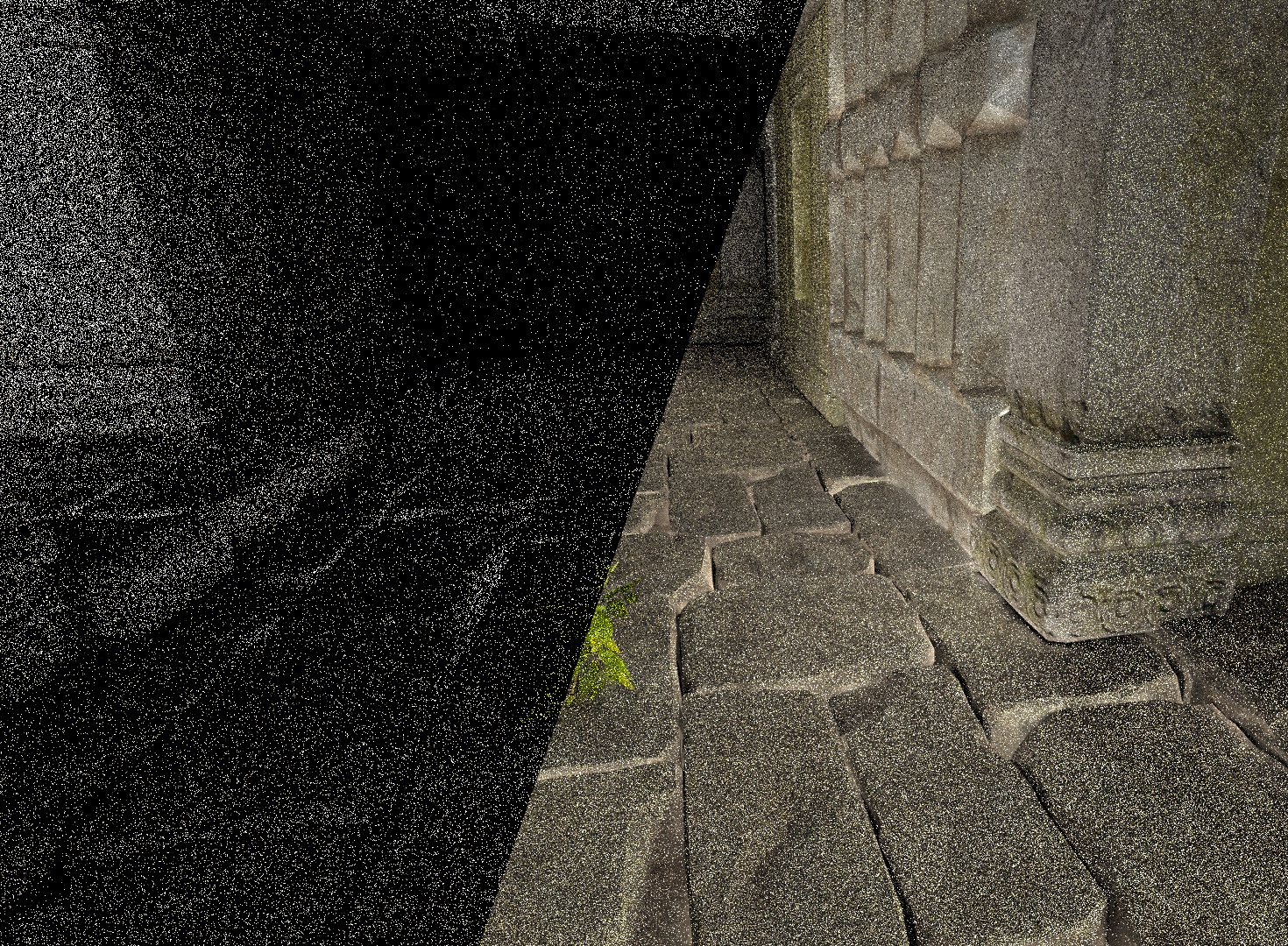}
	\caption{Indirect illumination by hashed path space filtering only at the second bounce:
	At 16 paths per pixel (left), the variance of the integrand is dramatically reduced (right).}
	\label{Fig:Indirect}
\end{figure}

While path space filtering dramatically reduces the noise at low sampling rates (see \cref{Fig:Rainforest}),
some noise is added back by spatial jittering.
Instead of selecting the first sufficiently diffuse vertex along a path from the camera,
path space filtering can be applied at any vertex.
For example, filtering at the second sufficiently diffuse vertex as shown in \cref{Fig:Indirect}
resembles final gathering or local passes \cite{PhotonMap:01}.
Further, it is obviously also possible to filter in several vertices along the path at the same time.
In fact, path space filtering trades variance reduction for controlled bias
and is orthogonal to other filtering techniques.
We therefore abstain from comparisons with these:
Temporal anti-aliasing and complimentary noise filters in screen space are appropriate to further reduce noise \cite{Mara:2017}.
A local smoothing filter \cite{Schied:2017} can even help reduce the error inherent in the approximation.

\section{Conclusion}
\label{Sec:Conclusion}

Relying on only a few synchronizations during accumulation,
path space filtering based on hashing scales on massively parallel hardware.
Both accumulation as well as queries run in constant time per vertex,
and neither the traversal nor the construction of a hierarchical spatial acceleration data structure is required.
Ray tracing hardware, accelerating ray intersection with the scene, is orthogonal to the presented method.
In fact, reasonable visual fidelity can often only be achieved
if ray tracing hardware is combined with such a powerful variance reduction technique
due to the time constraints for real-time simulations.

The simplistic algorithm overcomes many restrictions of screen space filtering,
does not require motion vectors,
and enables noise removal beyond the first intersection
including specular and transparent surfaces.

The hashing scheme still bears potential for improvement.
For example, important hashes could be excluded from eviction by reducing the resolution,
i.e. accumulating their contributions at a coarser level.
Other than selecting the resolution by the length of the path,
path differentials and variance may be used to determine the appropriate resolution.

Besides the classic applications of path space filtering \cite[Sec.3]{PathSpaceFiltering}
like multi-view rendering, spectral rendering, participating media,
and decoupling anti-aliasing from shading,
the adaptive hashing scheme can be used for photon mapping \cite{PhotonMap:01,Hachisuka:2009:SPPM}
and irradiance probes
in reinforcement learned importance sampling \cite{LightRL}
in combination with final gathering to store radiance probes.
Since the first publication of this work as a technical report
variants of the presented method has been applied to
improving the efficiency of \emph{ambient occlusion} \cite{GautronHashedAO}
and for an efficient implementation of reinforcement learning \cite{PantaleoniHashed}.

\section*{Acknowledgements}
\label{Sec:Acknowledgements}

The authors would like to thank Petrik Clarberg for profound discussions and comments.

\newcommand{\etalchar}[1]{$^{#1}$}

\end{document}